\documentclass[12pt,aps,floatfix,preprint,showpacs,amsmath,amssymb]{revtex4}
\usepackage{graphicx}
\usepackage{dcolumn}
\usepackage{bm}
\usepackage{epsfig,psfrag,amsmath,amssymb,float}
\input{epsf}
\begin{document}
\title{The Gap Function {\boldmath $\phi (k, \omega)$}
for a Two-leg t-J Ladder\\
and the Pairing Interaction}

\author{Didier~Poilblanc}
\email{didier.poilblanc@irsamc.ups-tlse.fr}
\affiliation{Laboratoire de Physique Th\'eorique CNRS-UMR5152,
Universit\'e Paul Sabatier, F-31062 Toulouse, France}

\author{D.~J.~Scalapino}
\email{djs@vulcan2.physics.ucsb.edu}
\affiliation{Department of Physics, University of California,
 Santa Barbara, CA 93106 USA}

\date{\today}

\begin{abstract}

The gap function $\phi(k, \omega)$, determined from a Lanczos calculation
for a doped 2-leg t-J ladder, is used to provide insight into the spatial
and temporal structure of the pairing interaction. It implies that this
interaction is a local near-neighbor coupling which is retarded. The onset
frequency of the interaction is set by the energy of an $S=1$
magnon-hole-pair and it is spread out
over a frequency region of order the bandwith.

\end{abstract}

\maketitle

The momentum and frequency dependence of the superconducting gap function
$\phi(k, \omega)$ provides information on the spatial and dynamic structure of
the pairing interaction. Traditionally, electron tunneling has been used to
explore the frequency dependence of the gap~\cite{MR69}.
Presently, angular resolved
photoemission spectroscopy \cite{DHS03} (ARPES) and scanning tunneling
microscopy \cite{Hof02, How03} (STM)
open the possibility of obtaining both $k$ and $\omega$ information about
the superconducting gap. However, approaches to extract this information on
$\phi(k, \omega)$ from such experiments are still being explored. As has
been previously discussed, a $k$ and $\omega$ dependent gap function for
finite t-J lattices can be obtained using Lanczos exact diagonalization
\cite{PS02, PSC04}.
Here, using results for $\phi (k, \omega)$ obtained for a doped 2-leg t-J
ladder, we explore what can be learned once data for
$\phi(k, \omega)$ becomes available.

The Hamiltonian for a 2-leg t-J ladder can be written as
\begin{eqnarray}
H & = & J_{\rm rung} \sum_i \left({\bf S}_{i1} \cdot {\bf S}_{i2} - \frac{1}{4}
\ n_{i1} n_{i2}\right) + J_{\rm leg} \sum_{i, \alpha} \left({\bf
S}_{i+1,\alpha} \cdot {\bf S}_{i, \alpha} - \frac{1}{4}\ n_{i+1,\alpha}
n_{i\alpha}\right)\nonumber\\
& - & t_{\rm rung} \sum_i \left(c^\dagger_{i, 1} c_{i,2} +
h.c.\right) - t_{\rm leg} \sum_{i, \alpha} \left(c^\dagger_{i+1,\alpha}
c_{i\alpha} + h.c.\right)
\label{one}
\end{eqnarray}
where $c_{i\alpha}$ are projected hole operators (spin indices are
omitted) and $\alpha = (1, 2)$ labels the two legs of the ladder.
We will consider the isotropic case in which $J_{\rm rung} =
J_{\rm leg} = J$ and $t_{\rm rung} = t_{\rm leg} = t$ and our
calculations will be carried out for a periodic $2\times 12$
ladder at $1/8$ and $1/6$ hole doping.

As previously discussed, a gap function can be extracted by combining
Lanczos results for the usual one-electron Green's function $G(k, \omega)$
with the Fourier transform of Gorkov's off-diagonal Green's function
\begin{equation}
F(k, t) = i \left\langle T c_{-k, -\sigma} (t/2) c_{k, \sigma}
(-t/2)\right\rangle \label{two}
\end{equation}
Here, for a finite system, this expectation value is taken between
the ground states for $N$ and $N-2$ particles and we choose the
phase of $F$ to be zero. We will take $N=22$, $N-2=20$
corresponding to an average filling $n\simeq \frac{21}{24}=0.875$.
The Dyson equations \cite{note1} relating $G$ and $F$ are
\begin{equation}
[Z(k, \omega)\, \omega - (\epsilon_k + X(k, \omega)]\, G(k, \omega) = 1 -
\phi (k, \omega)\, F(k, \omega)
\label{three}
\end{equation}
\begin{equation}
[Z(-k, -\omega)\, \omega - (\epsilon_k + X(-k, -\omega))]\, F(k, \omega)
= - \phi (k, \omega)\, G(k, \omega)
\label{four}
\end{equation}
with $Z$ and $X$ the usual Nambu self energies and $\phi(k,
\omega)$ the gap function. Then, using the even $(k\to -k)$ parity
of $Z(k, \omega)$, $X(k, \omega), \epsilon_k$ and $\phi(k,
\omega)$ along with the even $\omega$ dependence of $Z(k,
\omega)$, $X(k, \omega)$ and $\phi(k, \omega)$, we have
\begin{equation}
\phi (k, \omega) = \frac{F(k, \omega)}{F^2(k, \omega) + G(k, \omega)
\, G(k, -\omega)}\, .
\label{five}
\end{equation}
Alternatively, $Z(k, \omega)$ can be eliminated to obtain an expression for
the superconducting gap
\begin{equation}
\Delta(k, \omega) = \frac{\phi(k, \omega)}{Z(k, \omega)} = \frac{2\omega\,
F(k, \omega)}{G(k, \omega) - G(k, -\omega)}\, .
\label{six}
\end{equation}
\noindent From a numerical calculation of $G(k, \omega)$ and $F(k, \omega)$,
$\Delta(k, \omega)$ has been obtained for a 32-site, t-J cluster
\cite{PS02}.  This
extended previous work by Ohta {\it et.~al} \cite{OSEM94} who fit the spectral
weight $Im\, F(k, \omega)$ to a $d_{x^2-y^2}$ BCS-Bogoliubov quasiparticle
form in which the frequency dependence of the gap was neglected.

In contrast to the long range order of the superconducting ground state of a 2D
lattice, a 2-leg ladder exhibits power law pair field correlations
\cite{DR96, Hay95} which
decay as $x^{-1/\kappa_\rho}$. Here, $\kappa_\rho$ is the Luttinger liquid
parameter associated with the massless charge mode. This implies that for a
ladder of length $L$, the off-diagonal Green's function $F(k, \omega)$
decays \cite{CP02, Voi98, OP03}
as $(\xi/L)^{\frac{1}{2 \kappa_\rho}}$. Here,
the coherence length $\xi$ is
proportional to the inverse of the gap. For our doped ladder $L=12$ is
of order this coherence length so that we can probe the internal $k-\omega$
structure of a pair.

For J/t $=$ 0.4 and $1/8$ doping, the $k$-dependence of the zero
frequency gap $\phi(k_x, k_y, \omega=0)$ is plotted in Fig.~1 for
the bonding $(k_y=0)$ and antibonding $(k_y=\pi)$ bands. For comparison,
the solid curves correspond to $\phi_0\, (a\cos\, k_x-\cos\, k_y)$ with
$\phi_0=0.3$ and $a=0.8$. This d-wave-like $k$-dependence of the gap function is
similar to the behavior found in previous studies of both the t-J
\cite{OSEM94} and Hubbard ladders \cite{DS97}.  It implies
 that the spatial structure of the pairing
interaction is dominantly a near-neighbor interaction.

The frequency dependence of the real and imaginary parts of
$\phi(k, \omega)$ are shown in Fig.~2 for $k$ values which are
near the fermi surface of the bonding (red) and antibonding
(green) bands respectively. In general, the gap function $\phi (k,
\omega)$ is a complex frequency-dependent function
$\phi_1+i\phi_2$ with the imaginary part associated with dynamic
decay processes. For our doped 2-leg ladder with J/t $=$ 0.4, the
magnitudes of the zero frequency gaps $\Delta_0=\Delta(k_F, 0)$
obtained from eq.~\eqref{six}, for both the bonding and
antibonding fermi points, are of order 0.15t. The gap, $\Delta_0$,
is reduced from $\phi(k_F, 0)$ due to the renormalization factor
$Z$. The onset of the imaginary part of $\phi_2(k, \omega)$ seen
in Fig.~2(b) appears to occur somewhat below $3\Delta_0$. In
addition, the peaking in $\phi_2(k, \omega)$ and the rapid rise in
$\phi_1(k, \omega)$ as this onset frequency is approached suggest
that a particular excitation mode occurs at a frequency $\Omega$
such that $\omega=\Delta_0 + \Omega$ determines the onset of $\phi_2(k,
\omega)$.

The occurrence of an onset peak in $\phi_2(k, \omega)$ at
threshold and the short-range nature of the interaction involving
scattering of pairs from the bonding to antibonding band $(q_y\sim
\pi)$, imply that the pairing in this energy regime is mediated by
an $S=1$ channel. This follows from the form of the coherence
factor which varies as
$
\frac{1}{2}\, \left(1 \pm \frac{\Delta (p+q, \omega)\, \Delta (p,
\omega)}{E(p+q)\, E(p)}\right),
$
with the plus sign associated with the charge channel and the minus sign
with the $S=1$ spin channel.  For a ``d-wave-like'' gap with ${\bf q} =
(k_F\, ({\rm bonding}) - k_F\, ({\rm antibonding}), \pi)$ and $\omega$ near
threshold, the coherence factor goes to 1 for the spin channel and
vanishes for the charge channel.

A plot of the low-energy $S=1$ excitations for a doped ladder is
shown in Fig.~3(a). The solid diamonds show a collective $S=1$
bound magnon-hole-pair \cite{Poi00, Poi04} mode and the open
symbols the $S=1$ particle-hole continuum. A measure of the
spectral weight of the $S=1$ channel which couples to $\phi(k,
\omega)$ is given by the d-wave projection of the spin fluctuation
spectral weight
\begin{equation}
V_d (\omega) = \frac{1}{N^2}\ \sum_{k,k^\prime}
(\cos k_x-\cos k_y)
\, (\cos k^\prime_x - \cos k^\prime_y)\ S({\bf k}-{\bf k}^\prime, \omega)
\label{seven}
\end{equation}
with
\begin{equation}
S(q, \omega) = - \frac{1}{\pi}\ Im \left\langle S^\alpha_{-q}
\ \frac{1}{\omega + E_0 + i\eta-H}\ S^\alpha_q\right\rangle\, .
\label{eight}
\end{equation}
Here $S^\alpha_q$ is the Fourier transform of the $\alpha$ spin
component. A plot of $V_d (\omega)$ versus $\omega$ is shown in
Fig.~3(b). The peak at $\Omega \simeq 0.15t$ arises from the bound
magnon-hole-pair mode and the high frequency weight comes from the
particle-hole continuum. We believe that the onset of $\phi_2(k,
\omega)$ seen in Fig.~2 is due to a process in which a single
particle excitation with energy $\omega=\Omega_m+\Delta_0$ emits a
bound magnon-hole-pair and drops down to the gap edge while the
remaining $\phi_2(k, \omega) \not= 0$ region arises from a
coupling to the $S=1$ continuum.

In order to further characterize the dynamic nature of the pairing
interaction, we introduce
\begin{equation}
I(k, \Omega) = \frac{1}{\pi} \int^\Omega_0 d\omega^\prime\ \frac{\phi_2(k,
\omega^\prime)}{\omega^\prime}\, .
\label{nine}
\end{equation}
Then, since the gap function satisfies a dispersion relation, one has
\begin{equation}
\phi_1(k, 0) = I(k, \Omega \to \infty) + \phi_{\rm static} (k)\, .
\label{ten}
\end{equation}
Here, $\phi_{\rm static} (k)$ represents a non-retarded
contribution.  For example, if one were to make a mean-field approximation
in which
\[
J\left({\bf S}_i \cdot {\bf S}_j - \frac{1}{4}\, n_i n_j\right) \to -
J\left(\langle \Delta_{i\delta}\rangle\, \Delta^\dagger_{i\delta} +
h.c.\right)
\]
one would have an effective attractive pairing interaction which is
independent of frequency which would contribute to $\phi_{\rm static}$.
In Fig.~4(a), we have plotted $I(k,
\Omega)/\phi_1(k, 0)$ versus $\Omega$ and one sees that for J/t $=$ 0.4,
$I(k, \Omega)$ saturates at over 80\% of $\phi_1(k, 0)$.
This implies that the dominant part
of the pairing interaction comes from dynamic processes.
Also shown, Fig.~4(b), are similar
results for J/t$=$0.8. In this, unphysically large, J/t regime only 60\%
of the $\omega=0$ gap function is associated with a dynamic pairing
interaction.
This is similar to earlier results for the spatial structure of a pair
\cite{Poi94, WS97} in
which near-neighbor sites for the two holes making up a pair were favored
for large values of J/t. For physical values of J/t, next-near-neighbor
(diagonal) hole-hole occupation was favored. The diagonal structure arises
dynamically and reflects the retarded nature of the pairing
interaction for physical values of J/t.

In summary, knowledge of the $k$- and $\omega$-dependence of the gap
function provides information on the spatial and temporal structure of the
pairing interaction. Numerical solutions for the two-leg ladder show that
the pairing interaction has a short-range, near-neighbor form, leading to
momentum transfer processes which scatter pairs between the bonding
and antibonding states. The d-wave-like momentum dependence of the gap function
and the $\omega$ onset
of $\phi_2(k, \omega)$ imply that the pairing interaction in this energy
regime is mediated by the $S=1$ channel. This channel contains both a bound
magnon-pair state and a continuum of particle-hole excitations. We believe
that the magnon-pair mode is responsible for the onset behavior seen in
$\phi_2(k, \omega)$. The
dispersion relation for $\phi_1 (k, 0)$ shows that for physical values of
J/t, the dominant part of the interaction is dynamic with
contributions coming from both the magnon-hole-pair mode and
the particle-hole continuum.

\begin{acknowledgments}
DJS would like to acknowledge support under NSF Grant \#DMR02-11166.
DP thanks IDRIS (Orsay, France) for use of supercomputer facilities.
\end{acknowledgments}

\begin{figure}
\begin{center}
\epsfig{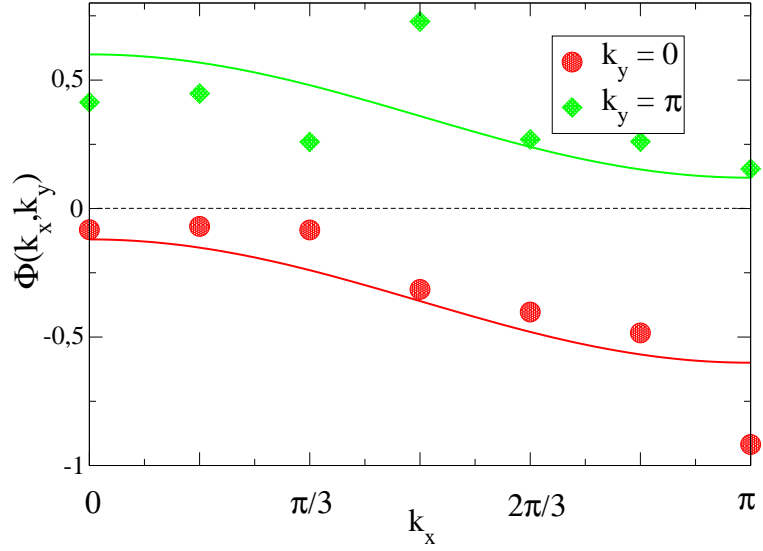} \caption{The $k_x$
dependence of  $\phi_1(k_x, k_y, \omega=0)$ for the bonding
$k_y=0$ (red) and antibonding $k_y=\pi$ (green) bands. For comparison, the solid
curves show $\phi_0\, (a\cos\, k_x-\cos\, k_y)$ with $\phi_0=0.3$ and $a=0.8$.}
\end{center}
\end{figure}

\begin{figure}
\begin{center}
\epsfig{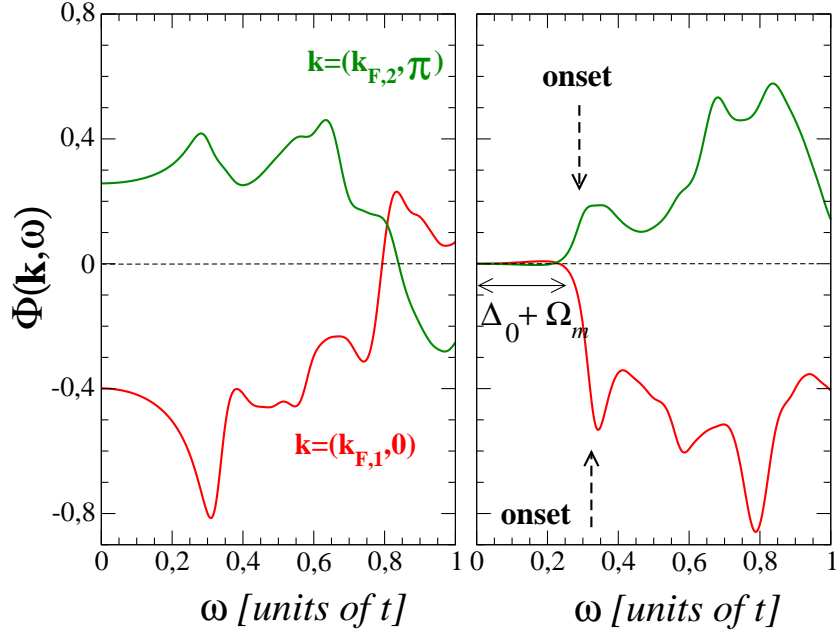}
 \caption{Real
(a) and imaginary (b) parts of the gap function $\phi(k, \omega)$
versus $\omega$ for a 2 $\times$ 12 ladder at values of $k$ near
the bonding (red) and antibonding (green) fermi momenta [$\sim
(\frac{2\pi}{3}, 0)$ and $\sim (\frac{\pi}{3}, \pi)$ respectively]
for J/t $=$ 0.4.  Here, $\Delta_0 \sim \Delta (k_{F}, \Delta_0)$
and $\Omega_m$ is the minimum energy for a bound magnon-hole-pair
excitation.}
\end{center}
\end{figure}

\begin{figure}
\begin{center}
\epsfig{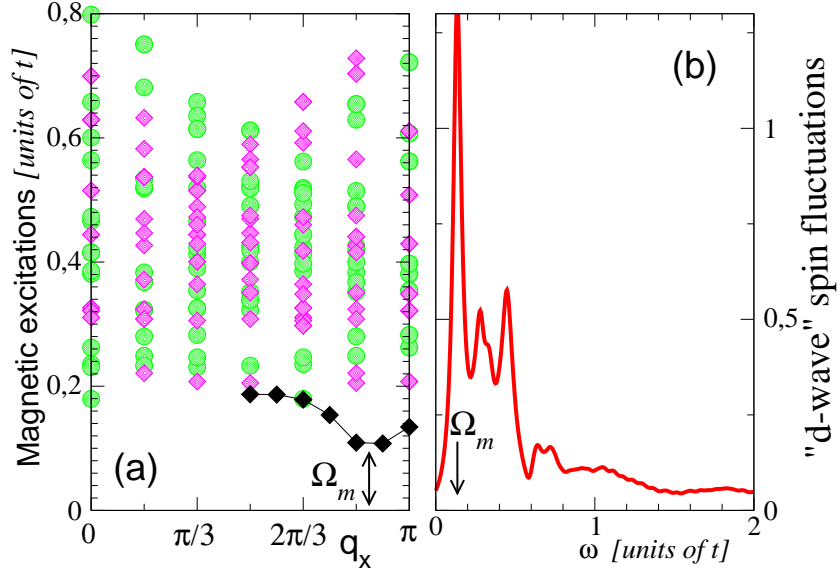}\caption{(a)
Low-energy electron-hole $S=1$ excitations for a $1/8$-doped J/t
$=$ 0.4 ladder for $q_y=0$ (circles) and $q_y=\pi$ (diamonds)
momenta. The solid diamonds denote a bound magnon-hole-pair
collective mode~\cite{PSC04}. (b) ``d-wave'' projection of the
spin fluctuation spectral weight with $\Omega_m$ the minimum
energy of the bound magnon-hole-pair collective mode ($1/6$
doping).}
\end{center}
\end{figure}

\begin{figure}
\begin{center}
\epsfig{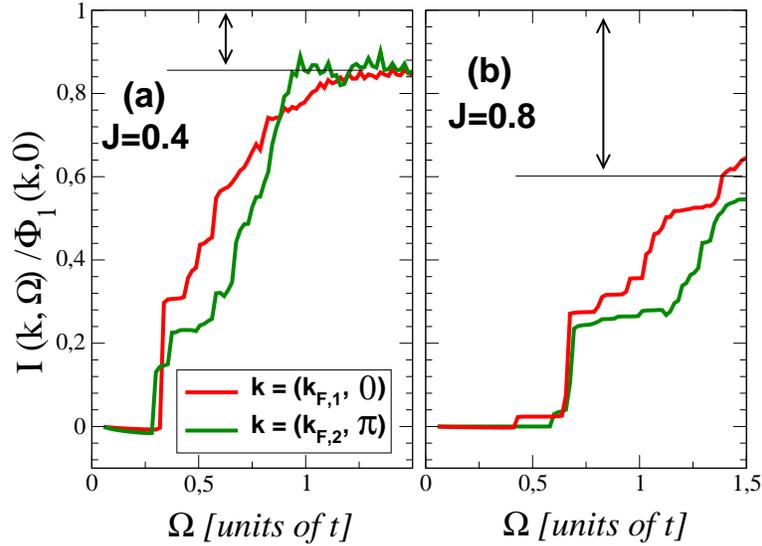}
\caption{The integrated contribution of $\phi_2(k, \omega)/\omega$
to $\phi_1(k, 0)$ for the bonding (red), antibonding (green)
$k$-values shown in Fig.~2.  (a) for J/t $=$ 0.4 and (b) for J/t
$=$ 0.8. The non-retarded contributions are shown by arrows.}
\end{center}
\end{figure}


\begin{thebibliography}{99}
\bibitem{note1} In general, the self-energy is decomposed as
$X(k,\omega)+\omega(1-Z(k,\omega))$. Although the 2-leg ladder is
a Luttinger liquid, the Green's functions for the finite $2\times
12$ lattice that we have studied can be parameterized in terms of
$Z(k, \omega)$, $X(k, \omega)$, and $\phi(k, \omega)$.
\end{thebibliography}
\end{document}